\documentclass[aps,prl,twocolumn,showpacs]{revtex4}
\usepackage{graphicx}
\usepackage{dcolumn}
\setcitestyle{super}

\begin{document}

\preprint{APS/123-QED}

\title{Hydrostatic-pressure tuning of magnetic, nonmagnetic and superconducting states in annealed Ca(Fe$_{1-x}$Co$_{x}$)$_{2}$As$_{2}$}
\author{E. Gati$^{1}$}
\author{S. K\"{o}hler$^{1}$}
\author{D. Guterding$^{1}$}
\author{B. Wolf$^{1}$}
\author{S. Kn\"{o}ner$^{1}$}
\author{S. Ran$^{2,3}$}
\author{S.L. Bud'ko$^{2,3}$}
\author{P.C. Canfield$^{2,3}$}
\author{M. Lang$^{1}$}

\address{$^{1}$Physikalisches Institut, J.W. Goethe-Universit\"{a}t
Frankfurt(M), SPP1458, D-60438 Frankfurt(M), Germany}
\address{$^{2}$Ames Laboratory, Iowa State University, Ames, Iowa 50011, USA}
\address{$^{3}$Department of Physics and Astronomy, Iowa State University, Ames, Iowa 50011, USA}

\date{\today}

\begin{abstract}
We report on measurements of the magnetic susceptibility and electrical resistance under He-gas pressure on single crystals of Ca(Fe$_{1-x}$Co$_{x}$)$_{2}$As$_{2}$. We find that for properly heat-treated crystals with modest Co-concentration, $x$ = 0.028, the salient ground states associated with iron-arsenide superconductors, i.e., orthorhombic/antiferromagnetic (o/afm), superconducting, and nonmagnetic collapsed-tetragonal (cT) states can be accessed all in one sample with reasonably small and truly hydrostatic pressure. This is possible owing to the extreme sensitivity of the o/afm (for $T \leq$ $T_{s,N}$) and superconducting ($T \leq T_c$) states against variation of pressure, disclosing pressure coefficients of d$T_{s,N}$/d$P$ = -(1100 $\pm$ 50) \,K/GPa and d$T_{c}$/d$P$ = -(60 $\pm$ 3)\,K/GPa, respectively. Systematic investigations of the various phase transitions and ground states via pressure tuning revealed no coexistence of bulk superconductivity (sc) with the o/afm state which we link to the strongly first-order character of the corresponding structural/magnetic transition in this compound. Our results, together with literature results, indicate that preserving fluctuations associated with the o/afm transition to low enough temperatures is vital for sc to form.
\end{abstract}


\maketitle

\begin{figure}[floatfix]
\begin{center}
  \includegraphics[width=0.9\columnwidth]{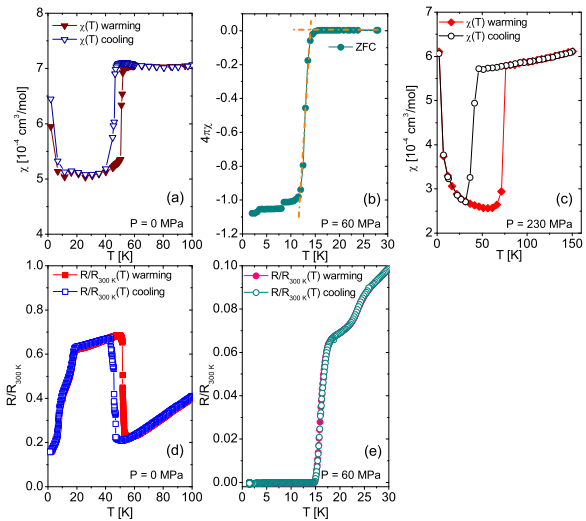}\\[-0.3cm]
  \caption{(Color online) Magnetic susceptibility $\chi$($T$) ((a), (b), (c)) and in-plane electrical resistance, normalized to the room-temperature value, $R$($T$)/$R_{300K}$ ((d), (e)), of Ca(Fe$_{1-x}$Co$_{x}$)$_{2}$As$_{2}$ with $x$ = 0.028/$T_{anneal}$ = 350$^{\circ}$C at $P$ = 0\,MPa ((a), (d)), 60\,MPa ((b), (e)) and 230\,MPa (c). $\chi$ data were taken in $B$ = 1\,T ((a), (c)) and 1\,mT (b) after zero-field-cooling (ZFC). Small step in $\chi$($P$ = 60\,MPa) (b) around 8\,K results from the solidification of $^{4}$He while step at 3\,K marks $T_{c}$ of a small In sample used as a manometer.} \label{fig:1}
\end{center}
\end{figure}

Among the various iron-arsenide-based superconductors, members of the $A$Fe$_{2}$As$_{2}$ ($A$ = Ba \cite{Rotter08a, Rotter08b, Ni08a}, Sr \cite{Yan08} and Ca \cite{Ni08b}) (122) family have become model systems for exploring the sc in this new class of superconductors. The 122 parent compounds do not manifest sc at ambient pressure ($P$) but rather undergo a phase transition from a high-temperature (high-$T$) tetragonal, paramagnetic state to a low-$T$ o/afm state. The o/afm phase below $T_{s,N}$ can be suppressed by chemical substitution \cite{Rotter08a, Sefat08, Ni10, Ren09, Thaler10} or pressure \cite{Colombier09, Matsubayashi09, Torikachvili08, Kotegawa09}, and sc develops. For the parent compounds, yet another phase, a nonmagnetic cT phase, has been observed at high pressure \cite{Torikachvili08, Kreyssig08, Goldman09, Uhoya10, Mittal11, Uhoya11}. Exploring the interplay of these various types of order has become a major theme of research in iron-arsenide materials. Of particular interest is whether sc coexists with afm order in the so-called "underdoped" areas of the phase diagram \cite{Pratt09, Rotter09, Aczel08, Goko09, Park09, Julien09} as this aspect is thought to hold the clue for discriminating unconventional $s^{+-}$ type of sc from conventional $s^{++}$ one \cite{Fernandes10a, Fernandes10b}.  \\
An important step towards clarifying these issues in a systematic and clean fashion was provided by recent investigations on Ca(Fe$_{1-x}$Co$_{x}$)$_{2}$As$_{2}$ single crystals \cite{Ran11, Ran12}. For these materials a postgrowth thermal treatment, involving an annealing/quenching temperature $T_{anneal}$, was established as another control parameter to systematically tune the ground state of the system \cite{Ran11, Ran12}. Since $T_{anneal}$ determines the size and nature of the precipitates, and by this the amount of strain built up in the materials, it was suggested that $T_{anneal}$ mimics the effect of pressure \cite{Ran11}.\\
In this Letter we provide evidence for such a $P$-$T_{anneal}$ analogy by demonstrating that for properly heat-treated Ca(Fe$_{1-x}$Co$_{x}$)$_{2}$As$_{2}$ single crystals (e.g. with $x \sim$ 0.028), the salient ground states associated with iron-arsenide superconductors, i.e., o/afm, superconducting, and nonmagnetic cT states can be accessed all in one sample by applying very modest, truly hydrostatic (He-gas), pressure. This is made possible owing to the extreme sensitivity of the various types of order against pressure variations, disclosing extraordinarily large pressure coefficients. Through hydrostatic-pressure tuning, we were able to systematically study the various ground states and their mutual interplay with very fine resolution on the pressure axis. Our results, together with literature results \cite{Fernandes10a}, give clear indications under which conditions sc does or does not coexist with structural/afm order in the 122 family. \\
Measurements of the magnetic susceptibility were conducted by using a SQUID magnetometer (Quantum Design MPMS). The susceptibility data have been corrected for the contribution of the sample holder, including the pressure cell, which was determined independently. The electrical resistance was measured in a four-terminal configuration by employing a Linear Research (LR700) bridge. For both experiments CuBe cells \cite{Pressure} connected to He-gas compressors \cite{Pressure} were used for finite-$P$ measurements.  An In \cite{Jennings58}(InSb \cite{Kraak84}) sample was used for the SQUID (resistance) measurements for an \textit{in situ} determination of the pressure. The use of $^{4}$He as a pressure-transmitting medium ensures truly hydrostatic-pressure conditions as long as $^{4}$He is in the liquid phase -- an aspect which is of particular importance for the present experiments, given the high sensitivity of CaFe$_{2}$As$_{2}$ to nonhydrostatic-pressure conditions \cite{Yu09, Canfield09}. The single crystals of Ca(Fe$_{1-x}$Co$_{x}$)$_{2}$As$_{2}$ were grown out of an FeAs flux, see ref.\,\cite{Ran12} for growth and annealing details.\\
Figure \ref{fig:1} shows data of the magnetic susceptibility, $\chi$($T$), and normalized electrical resistance, $R$($T$)/$R_{300K}$, of single crystalline Ca(Fe$_{1-x}$Co$_{x}$)$_{2}$As$_{2}$ with $x$ = 0.028 and $T_{anneal}$ = 350$^{\circ}$C for a selection of pressure values. The data reveal distinctly different types of anomalies which are found to be representative for three distinct pressure ranges. At low pressure values, $P \leq$ 32\,MPa, represented by the $P$ = 0 data (Fig.\,\ref{fig:1}(a) and (d)), jump-like changes were observed in both quantities accompanied by a distinct hysteresis upon cooling and warming. The features observed here, i.e., the sharp and modest decrease in $\chi$ by about 2$\times$10$^{-4}$cm$^{3}$mol$^{-1}$ accompanied by a sizable discontinuous increase in $R/R_{300K}$ on cooling, are consistent with the $P$ = 0 results reported in ref.\,\cite{Ran12} on crystals with the same x and $T_{anneal}$. In this work, the jumps have been assigned to the strongly first-order transition from the high-$T$ tetragonal, paramagnetic state to the low-$T$ o/afm state. For the determination of the transition temperature, we refer to $\chi$ as a thermodynamic probe and assign the transition temperature to the point of maximal change in $\chi$($T$). Applying this criterion yields $T_{s,N}^{cool}$ = 46.3\,K and $T_{s,N}^{warm}$  = 51\,K for the transition measured upon cooling and warming, respectively. The sudden drop in $R$($T$) at around 20\,K, accompanied by a tiny feature in $\chi$, indicates the onset of filamentary sc, in accordance with literature \cite{Ran12}.\\
At intermediate pressure values, 32\,MPa $\leq P \leq$ 180\,MPa, represented by the $P$ = 60\,MPa data (Fig.\,\ref{fig:1}(b) and (e)), a large diamagnetic signal in $\chi$ accompanied by zero resistance is observed. Measurements after zero-field-cooling (ZFC) reveal a shielding signal corresponding to 100\% diamagnetism. By using the crossing point of linear extrapolations from below and above the onset of diamagnetism, a superconducting transition temperature of $T^{\chi}_{c}$ = 13.8\,K is inferred, which roughly coincides with the temperature of zero resistance of $T^{R}_{c}$ = 14.9\,K. Here too, we refer to the thermodynamic quantity $\chi$ for assigning the transition temperature. The broadened step-like reduction in $R$ below about 25\,K suggests some filamentary sc with higher $T_c$. Very similar observations, i.e., a full shielding signal accompanied by zero resistance, were made in ref.\,\cite{Ran12} on a sample with $x$ = 0.033 and $T_{anneal}$ = 350$^{\circ}$C, where the bulk character of sc was proven by specific heat measurements \cite{Budko09}.\\
At higher pressures, $P \geq$ 210\,MPa, represented by the $P$ = 230\,MPa data (Fig.\,\ref{fig:1}(c)), no further sc is observed. Instead, $\chi$($T$) shows a sharper drop, of about 3$\times$10$^{-4}$cm$^{3}$mol$^{-1}$ upon cooling, and an even more pronounced hysteresis than the low-$P$ features associated with $T_{s,N}$. Both the enhanced jump size and its positive pressure dependence distinguish this transition from the one at $T_{s,N}$, characterized by a huge negative pressure coefficient, see below. The phenomenology observed here is identical to that found for non-substituted CaFe$_{2}$As$_{2}$ without heat treatment or Ca(Fe$_{1-x}$Co$_{x}$)$_{2}$As$_{2}$ with $x$ $\geq$ 0.01 and $T_{anneal} \geq$ 600$^{\circ}$C \cite{Ran12} where structural investigations have identified this feature as the transition into the low-$T$, nonmagnetic, cT phase. Since large lattice deformations accompany this phase transition, often leading to cracks within the sample and/or the loss of electrical contacts, no resistance data could be obtained below $T_{cT}$ in the present study, consistent with the observations in ref.\,\cite{Ran12}. From the positions of the largest drops in $\chi$($T$) we derive the corresponding transition temperatures of $T^{cool}_{cT}$ = 39\,K and $T^{warm}_{cT}$ = 73.6\,K. Since $T_{cT}$ is accompanied by a pronounced hysteresis as a function of $P$ at fixed $T$ \cite{Goldman09}, the temperature sweeps reported here have been performed in a sequence with increasing pressure.

\begin{figure}[floatfix]
\begin{center}
  \includegraphics[width=0.9\columnwidth]{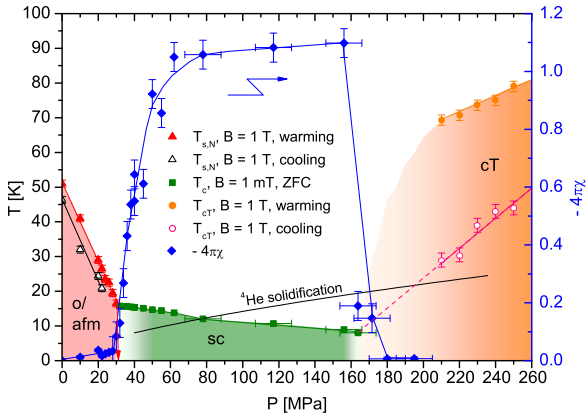}\\[-0.3cm]
   \caption{(Color online) $T$-$P$ phase diagram of single crystalline Ca(Fe$_{1-x}$Co$_{x}$)$_{2}$As$_{2}$ with $x$ = 0.028/$T_{anneal}$ = 350$^{\circ}$C inferred from $\chi$($T$) data. Filled (open) up-triangles correspond to the transition into the low-$T$ o/afm phase at $T_{s,N}$. Filled squares represent transition into the sc phase at $T_{c}$ inferred from ZFC measurements. For those $T_{c}$ values determined below the solidification line of $^{4}$He (black solid line), the $P$ values have been corrected by a factor 0.78 to account for the $P$ drop accompanying solidification. Closed diamonds indicate the size of the diamagnetic signal (without correcting for demagnetization effects) (right scale) in units of -4$\pi$, where -4$\pi$ corresponds to 100\% shielding volume. Filled (open) circles correspond to transition into the low-$T$ cT phase. }

 \label{fig:2}
\end{center}
\end{figure}

\begin{figure}[floatfix]
\begin{center}
  \includegraphics[width=0.9\columnwidth]{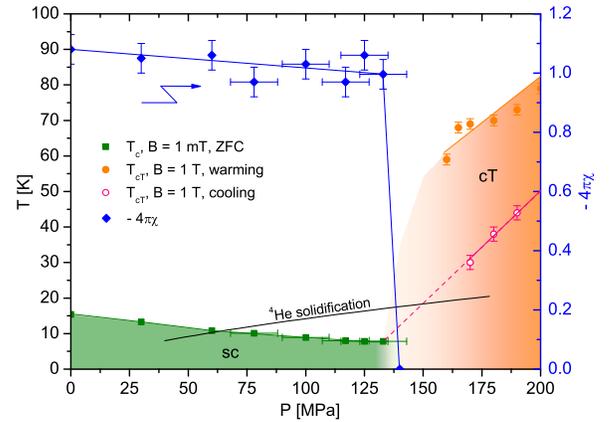}\\[-0.3cm]
  \caption{(Color online) $T$-$P$ phase diagram of Ca(Fe$_{1-x}$Co$_{x}$)$_{2}$As$_{2}$ with $x$ = 0.029 and $T_{anneal}$ = 400$^{\circ}$C inferred from $\chi$($T$) measurements. Filled squares represent $T_{c}$ values (left scale) inferred from ZFC measurements and filled diamonds the corresponding diamagnetic shielding volume (right scale). Filled (open) circles correspond to $T_{cT}$ as inferred from measurements upon warming (cooling). Note that in a cooling and subsequent warming run at $P$ = 160\,MPa a measurable $\Delta\chi$ at $T_{cT}$ was revealed only upon warming. Black solid line indicates the solidification of $^{4}$He.} \label{fig:3}
\end{center}
\end{figure}

After having identified the nature of the various anomalies observed in $\chi$($T$) and $R$($T$) measurements and having determined criteria for inferring phase transition temperatures, we compile the data for $x$ = 0.028 and $T_{anneal}$ = 350$^{\circ}$C in a $T$-$P$ phase diagram in Fig.\,\ref{fig:2}. The figure highlights the extraordinarily high sensitivity of the o/afm transition to pressure: upon increasing pressure $T_{s,N}$ becomes reduced in a linear fashion from $T_{s,N}$ = 51\,K (0\,MPa) to 29.5\,K (20\,MPa) and 16.5\,K (30\,MPa), corresponding to an unprecedentedly large pressure dependence of d$T_{s,N}$/d$P$ = -(1100 $\pm$ 50) \,K/GPa. The strongly hysteretic behavior revealed in the $\chi$($T$) and $R$($T$)/$R_{300K}$ measurements demonstrates that the o/afm transition remains first order within this pressure range. At the same time we observe the occurrence of some filamentary sc with $T_c \simeq$ 15\,K the volume fraction of which gradually grows from 0 (0\,MPa) to about 1\% (10\,MPa) and 3\% (28\,MPa). Upon further increasing the pressure to $P$ = 32\,MPa, however, no discontinuous changes accompanied by hysteretic behavior down to 2\,K, (1.6\,K), the lowest temperature in the magnetic (resistance) measurements, were found. This suggests that at this pressure level no phase transition line into the o/afm phase has been crossed in the $T$ range investigated. Instead, the data show zero resistance both in cooling and warming runs and a superconducting shielding volume which starts growing rapidly, reaching about 12\% (60\%) at $P$ = 32\,MPa (40\,MPa). A full (100\%) diamagnetic shielding volume, is revealed for $P$ above about 50$-$60\,MPa up to 156\,MPa. In this pressure range $T_c$ shows, to a good approximation, a linear reduction with $P$ from 13.8\,K (60\,MPa) to 9\,K (156\,MPa). This corresponds to a pressure coefficient of d$T_{c}$/d$P$ = -(60 $\pm$ 3)\,K/GPa, again exceptionally large.\\
By allowing for a weak distribution of strain within the sample, still compatible with our observation of sharp features at the o/afm transition, various scenarios are possible to rationalize these observations. On the one hand, the data would be compatible with the existence of a critical point ($T^{crit}$, $P^{crit}$) in the $T$-$P$ phase diagram, where the o/afm and the sc phase transition lines meet. The first-order character of the o/afm transition line would then imply a tricritical point, see, e.g.\,ref.\,\cite{Yip91}, either at finite $T^{crit}$ or at $T^{crit}$ = 0. On the other hand, the gradual growth of the superconducting volume fraction to 100\% over a finite pressure range above 30\,MPa may indicate that a small but finite gap exists on the $P$ axis, separating the critical pressure ($P_{c}^{o/afm} <$ 32\,MPa) where $T_{s,N}$ has dropped to zero, from the appearance of sc with 100\% shielding volume at $P > P_{c}^{o/afm}$.\\
Upon further increasing $P$ to above 156\,MPa, we observe, within a very narrow pressure range of 156\,MPa $\leq P \leq$ 180\,MPa, a sudden drop of the shielding signal to zero, i.e.\,, a complete loss of sc. A further increase in pressure to 210\,MPa is necessary to reveal first weak indications for the transition at $T_{cT}$ in our $\chi$($T$) measurements. The jump size observed here of $\Delta\chi \simeq$ -(1$-$1.5)$\times$10$^{-4}$cm$^{3}$/mol is a factor 2--3 smaller than the signatures found at $P\geq$ 230\,MPa, cf.\,Fig.\,\ref{fig:1}(c). Whereas $\Delta\chi$ stays practically constant for 220\,MPa $\leq P \leq$ 260\,MPa, $T_{cT}$ grows almost linearly with pressure at a rate of d$T_{cT}^{cool}$/d$P$ = +(420 $\pm$ 70)\,K/GPa. Note that the apparent strong reduction of the jump size for $P \leq$ 210\,MPa and the limitations in the accessible lowest temperature entailed by the solidification of $^{4}$He, cf.\,the solid line in Fig.\,\ref{fig:2}, make it impossible to track the cT phase transition line towards lower pressure values.\\
The progression of the transition temperature $T_{cT}$ with pressure shown in Fig.\,\ref{fig:2} suggests a close connection between the occurrence of the cT phase and the disappearance of sc: a linear extrapolation of the $T_{cT}^{cool}$ line towards lower pressure, cf.\,the broken line in Fig.\,\ref{fig:2}, truncates the $T_{c}$ line around the critical pressure $P_{c}^{sc}$ ($\sim$ 165\,MPa) above which sc disappears. We anticipate that the structural changes, and the accompanying magnetic signatures, at $T_{cT}$ become considerably reduced for $P \leq$ 210\,MPa. Yet, the structural changes are strong enough to suppress sc. To substantiate this hypothesis, we have carried out an analogous pressure study on another crystal with different $x$ and $T_{anneal}$, cf.\,Fig.\,\ref{fig:3}. According to ref.\,\cite{Ran12}, both a slight increase in $x$ as well as an enhancement of $T_{anneal}$ lead to a suppression of the o/afm phase and the emergence of sc at ambient pressure. Thus, for these crystals one may expect to observe the $P$-induced change from sc to the cT phase already at smaller pressure values. Figure \ref{fig:3} shows the results on a Ca(Fe$_{1-x}$Co$_{x}$)$_{2}$As$_{2}$ crystal with $x$ = 0.029 and $T_{anneal}$ = 400$^{\circ}$C. At $P$ = 0 the system shows a sc ground state with $T_{c}$ = 15.4\,K and full diamagnetic shielding, consistent with literature results \cite{Ran12}. Upon increasing the pressure to 133\,MPa, $T_{c}$ is reduced to 7.8\,K while the shielding signal stays essentially constant. Note that the corresponding pressure coefficient of d$T_{c}$/d$P$ = -(60 $\pm$ 3)\,K/GPa is identical to that obtained for the x = 0.028 and $T_{anneal}$ = 350$^{\circ}$C sample. Further similarities to the latter sample include the abrupt disappearance of sc within a very narrow pressure window, here 130\,MPa$-$140\,MPa, and the observation of magnetic signatures of the cT phase transition at somewhat higher pressures. Here too, the tracking of the magnetic signatures towards lower pressure is hampered by the strong reduction of the signature in $\chi$, here for $P \leq$ 170\,MPa, and the limitations set by the solidification of $^{4}$He. Yet, the available $T_{cT}$ data show the same characteristics as revealed for the x = 0.028 and $T_{anneal}$ = 350$^{\circ}$C sample, i.e.\,, a linear extrapolation of the $T_{cT}^{cool}$($P$), cf.\,the broken line in Fig.\,\ref{fig:3}, truncates sc.\\

Below we summarize the salient results of our pressure studies, indicate open issues, and discuss the implications of our findings for understanding sc in the 122 family.\\
(1) We have shown that carefully substituted and annealed samples of Ca(Fe$_{1-x}$Co$_{x}$)$_{2}$As$_{2}$, with suppressed $T_{s,N}$ values, as, e.g., $x$ = 0.028 and $T_{anneal}$ = 350$^{\circ}$C, represent a unique case where the salient states associated with iron-based superconductivity, i.e., o/afm, superconducting and cT phases, can be accessed all in a single sample by applying very modest, hydrostatic pressure. (2) At zero- and low-$P$ values, the system with x = 0.028 exhibits a strongly first-order o/afm transition which becomes rapidly suppressed with pressure. (3) The transition at $T_{s,N}$ remains first order up to at least 30\,MPa, the highest pressure value where clear signatures of $T_{s,N}$ could be revealed by susceptibility and resistance measurements. The data suggest the existence of a critical pressure $P^{o/afm}_{c} \simeq$ 32\,MPa where the o/afm phase is suppressed to zero, although other scenarios, such as a finite-$T$ or zero-temperature tricritical point, cannot be ruled out at present. (4) The application of $P \geq$ 50$-$60\,MPa (i.e., $P > P^{o/afm}_{c}$) is necessary to stabilize bulk sc, demonstrating that (5) for the present material there is no coexistence of sc with the o/afm phase. (6) Upon increasing the pressure to values somewhat above the critical pressure $P^{sc}_{c}$ where sc disappears abruptly, the cT phase can be detected via its magnetic signature, and further stabilized with pressure, implying that there is no coexistence between sc and the cT phase either. The data suggest that the cT phase extends down to lower pressures and truncates sc at $P$ = $P_{c}^{sc}$. Furthermore, (7) we provide evidence for the existence of a $P$-$T_{anneal}$ analogy for the present materials, indicating that here $T_{anneal}$ mimics the effect of pressure as suggested previously \cite{Ran11}. In fact, the various phase transition temperatures revealed for Ca(Fe$_{1-x}$Co$_{x}$)$_{2}$As$_{2}$ with x = 0.028/$T_{anneal}$ = 350$^{\circ}$C and 0.029/400$^{\circ}$C in the present pressure studies and those obtained from an x = 0.028 sample treated with varying $T_{anneal}$ \cite{Ran12}, can be combined in a composite, unified phase diagram. By using the conversion $\Delta T_{anneal}$ = 100$^{\circ}$C $\equiv$ 84.6\,MPa an almost perfect matching is obtained for both the $T_{o/afm}$ and $T_{c}$ lines for the various samples, while some sample-to-sample variations become apparent for the $T_{cT}$ line. The latter observation is likely to be caused by the temperature/pressure history dependence of $T_{cT}$ as a consequence of the extraordinarily large lattice deformations accompanying this transition. (8) The pressure coefficients of the various phase transitions lines revealed here of d$T_{s,N}$/d$P$ = -(1100 $\pm$ 50) \,K/GPa, d$T_{c}$/d$P$ = -(60 $\pm$ 3)\,K/GPa and d$T_{cT}^{cool}$/d$P$ = +(420 $\pm$ 70)\,K/GPa all are exceptionally large, by far the largest among all iron-based superconductors \cite{Chu09, Sefat11}. This illustrates how close to the edge of stability the parent compound CaFe$_{2}$As$_{2}$ actually is.\\
From these observations, together with literature results \cite{Fernandes10a}, some important conclusions can be drawn as for the interplay of sc with the nearby structural and afm orders that form in the 122 family. Most importantly, given the microscopic coexistence of competing superconducting and o/afm phases, well-established for Ba(Fe$_{1-x}$Co$_{x}$)$_{2}$As$_{2}$ \cite{Pratt09}, where the transition at $T_{s,N}$ is of second order, we link the non-coexistence in the present case to the strongly first-order character of the $T_{s,N}$ line. This finding, together with the absence of sc in the nonmagnetic cT phase, clearly indicate that preserving fluctuations associated with the o/afm transition to low enough temperatures is vital for sc to form here. We speculate that in the present first-order situation, the competition between sc and the o/afm order manifests itself in a separation of the two phases, i.e., a sudden drop of the $T_{s,N}$ line preceding the formation of sc at higher pressures, consistent with the experimental observations.\\ Finally we point out that the present results on Ca(Fe$_{1-x}$Co$_{x}$)$_{2}$As$_{2}$ hold great promise for further studies both as a function of temperature at $P$ = const. but also as a function of pressure at constant $T$, for a systematic investigation of the role of structural/afm orders and their fluctuations for sc in the iron-arsenide-based superconductors.\\

\begin{acknowledgments}
 Work done at Ames Lab (S.R., S.L.B., P.C.C.) was supported by the U.S. Department of Energy, Office of Basic Energy Science, Division of Materials Sciences and Engineering. Ames Laboratory is operated for the U.S. Department of Energy by Iowa State University under Contract No. DE-AC02-07CH11358. M.L. acknowledges fruitful discussions with P. Thalmeier.
\end{acknowledgments}

\end{document}